\documentclass[journal,final]{IEEEtran} % double

\usepackage{cite}

\ifCLASSINFOpdf
 \usepackage[pdftex]{graphicx}
 \graphicspath{{../pdf/}{../jpeg/}}
 \DeclareGraphicsExtensions{.pdf,.jpeg,.png}
\else
   \usepackage[dvips]{graphicx}
  \graphicspath{{./Figures/}}
  \DeclareGraphicsExtensions{.eps}
 \fi

\usepackage[cmex10]{amsmath}
\usepackage{amssymb}

\usepackage{epstopdf}
\usepackage{color}

\newtheorem{lemma}{Lemma}

\def\Kmax{K_{\text{max}}}
\def\x{\mathbf{x}}
\def\y{\mathbf{y}}
\def\res{\mathbf{r}}
\def\AOMP{{A$^\star$OMP}}
\def\AOMPK{{\AOMP$_K$}}
\def\AOMPe{{\AOMP$_e$}}

\def\Astar{A$^\star$}
\def\MMPKDF{MMP$_K$-DF}
\def\MMPeDF{MMP$_e$-DF}

\begin{document}

\title{Comments On ``Multipath Matching Pursuit''\\ by Kwon, Wang and Shim}
\author{Nazim~Burak~Karahanoglu and Hakan~Erdogan %
\thanks{
Nazim~Burak~Karahanoglu is with The Informatics and Information Security Research Center, The Scientific and Technological Research Council of Turkey (TUBITAK BILGEM), Kocaeli, Turkey. (email: burak.karahanoglu@tubitak.gov.tr)

Hakan~Erdogan is  with the Faculty of Engineering and Natural Sciences, Sabanci University, Istanbul, Turkey. (email: haerdogan@sabanciuniv.edu)

}%
}

\maketitle

\begin{abstract}

Straightforward combination of tree search with matching pursuits, which was suggested in 2001 by Cotter and Rao, and then later developed by some other authors, has been revisited recently as multipath matching pursuit (MMP).
In this comment, we would like to point out some major issues regarding this publication.
First, the idea behind MMP is not novel, and the related literature has not been properly referenced.
MMP has not been compared to closely related algorithms such as \Astar orthogonal matching pursuit ({\AOMP}).
The theoretical analyses do ignore the pruning strategies applied by the authors in practice.
All these issues have the potential to mislead the reader and lead to misinterpretation of the results.
With this short paper, we intend to clarify the relation of MMP to existing literature in the area and compare its performance with {\AOMP}.

\end{abstract}

\section{Introduction}

\IEEEPARstart{T}{he} idea of straightforward combination of  tree search and matching pursuits has been revisited recently as multipath matching pursuit (MMP) \cite{MMP2014}.
Despite MMP has been discussed in terms of both theoretical and empirical aspects, we believe that \cite{MMP2014} lacks some vital aspects including novelty and relations to prior work.
The absence of these issues can easily be misleading and can lead to misrepresentation of the algorithm in relation to the existing literature.
Consequently, we find it very important to clarify these issues, prevent potential misunderstanding and set the literary record straight for this family of algorithms.
With this motivation, the following sections of this comment address three main issues which are listed below:

\begin{enumerate}
\item \textbf{Novelty:} Straightforward combination of tree search with matching pursuit  algorithms as in MMP is  not novel. Cotter and Rao have already presented this idea in 2001 \cite{Cotter:TSBOMP}, however their work is not referenced in \cite{MMP2014}.
\item \textbf{Relations to existing sophisticated search techniques:}
MMP is closely related to {\Astar} orthogonal matching pursuit ({\AOMP}) \cite{Karahanoglu:AOMPfull, Karahanoglu:ImprovingAOMP} which performs a more sophisticated tree search via dynamic path selection.
% \cite{MMP2014} is missing a comparison of MMP to {\AOMP} which should have been presented for a fair evaluation.
MMP should have been compared to {\AOMP} in the first place, however this is missing in \cite{MMP2014}.
We  present empirical results, according to which {\AOMP} yields clearly higher recovery accuracy than MMP.
\item \textbf{Validity of theoretical analysis:}
 The recovery guarantees presented in \cite{MMP2014} ignore pruning which the authors incorporate for tractability in practice.
Though such limitations may be acceptable, their relations to the theory should be discussed in order to prevent misinterpretation.
In addition, the theoretical guarantees of {\AOMP} in \cite{Karahanoglu:ImprovingAOMP} are also valid for MMP and provide a looser condition for noise-free recovery of sparse signals.

% \cite{MMP2014} provides general recovery guarantees for MMP.
% However, the presented theoretical analysis ignores pruning which the authors incorporate for tractability in practice.
% Though such limitations may be accepted, their relations to the theory should be clarified in order to prevent misinterpretation of the analytical findings.
% In addition, the theoretical analysis of {\AOMP} in \cite{Karahanoglu:ImprovingAOMP} is also valid for MMP and provides a looser condition for noise-free recovery of sparse signals.

\end{enumerate}

\section{MMP is not Novel}
\label{sec:Novelty}

While original matching pursuit algorithms \cite{Mallat:MP, Pati:OMP} are greedy and decide on a single index at each iteration, a possible extension considers multiple alternative indices at each iteration and performs a tree search among alternative index sets to extract a final support set.
This idea has been suggested by Cotter and Rao \cite{Cotter:TSBOMP} in 2001.
% Straightforward combination of the tree search with matching pursuit type algorithms has been suggested by Cotter and Rao \cite{Cotter:TSBOMP} in 2001.
They have developed two strategies for parsing a search tree with branching factor $K$.
Among these,  MP:M-L is based on a breadth-first strategy.
It proceeds level by level, exploring $K$ children of each leaf and keeping the best $M$ nodes at the next level.
When the specified depth is reached, the path with the smallest residual is returned.
MP:K follows a depth-first nature processing the paths sequentially.
It explores a complete path and terminates if the residual is small enough.
Otherwise, the tree is backtracked and  other candidates are explored one by one until the solution is found.
Some modifications to increase the efficiency and tractability of these methods have later been suggested in \cite{Karabulut:FlexTreeSearch} and \cite{Schnitter:FBMP}.

\cite{MMP2014} recalls the idea developed by Cotter and Rao in \cite{Cotter:TSBOMP} under the name multipath matching pursuit.
As in \cite{Cotter:TSBOMP}, breadth-first (MMP-BF) and depth-first (MMP-DF) strategies are proposed.
MMP-BF is equivalent to MP:M-L\footnote{Though the definition of MMP-BF in \cite{MMP2014} skips pruning of leaf nodes at each iteration, the authors still employ this strategy in practice for tractability.}.
Similarly, MMP-DF is equivalent to MP:K except that MMP-DF places an upper limit on the number of explored paths.
Hence, neither the straightforward combination of tree search with matching pursuits nor the developed algorithms are new in \cite{MMP2014}.
Despite this fact, the authors do not credit Cotter and Rao or their successors.

% Similarly, MMP-DF is equivalent to MP:K except that \cite{MMP2014}  limits the number of explored paths.

% MMP is based on a trivial combination of tree search with matching pursuit algorithms.
% It explores a search tree with branching factor $L$.
% \cite{MMP2014} develops two variants of MMP.
% Breadth-first (MMP-BF) processes all leaves at a certain depth at once by exploring $L$ children of each leaf and then proceeds to the next depth.
% When the tree depth becomes $K$, the path with the lowest residual is returned.
% Depth-first (MMP-DF) follows a predefined order to explore paths sequentially.
% The algorithm first explores a complete path up to maximum depth $K$.
% If this path does not yield the desired solution, the tree is backtracked and  other candidates are explored sequentially until the solution is found.
% Since these strategies are expected to open too many nodes in practice, the authors employ some pruning strategies to ensure tractability \cite{MMP2014}.
% For MMP-BF, they explore only 50 candidates at each iteration. For MMP-DF, only the first $N_{\text{max}}=50$ paths are explored.

% This formulation of MMP is not novel in contrast to the claim in \cite{MMP2014}.
% The same idea has been proposed by Cotter and Rao in 2001 \cite{Cotter:TSBOMP}.
% Cotter and Rao propose two variants of their approach.

\section{Relations of MMP to {\AOMP}}
\label{sec:AOMP}
\subsection{From Straightforward to Sophisticated Tree Search}

As mentioned above, MMP belongs to a group of algorithms which are based on straightforward combinations of tree search with matching pursuits.
On the other hand, {\AOMP} algorithm \cite{Karahanoglu:AOMPfull, Karahanoglu:ImprovingAOMP} has been suggested before MMP in 2012 in order to incorporate more sophisticated search heuristics into this framework.
Based on a  combination of {\Astar} search with OMP, {\AOMP} employs dynamic path selection techniques.
The path selection mechanism of {\AOMP} enables comparison of candidates with different number of nonzero indices via adaptive cost models based on dedicated auxiliary functions and residual energy.
This allows for adaptive selection of promising candidates on-the-fly in contrast to the predefined order of MMP.
This more sophisticated strategy promises the potential to guide the search in an intelligent manner, and improve the recovery accuracy.
The enhanced recovery performance of {\AOMP} has been demonstrated via rich simulations including recovery probabilities, phase transitions and image recovery examples  in \cite{Karahanoglu:AOMPfull} and \cite{Karahanoglu:ImprovingAOMP}.
These results reveal that {\AOMP} is able to outperform the mainstream algorithms in the field in many scenarios in terms of recovery accuracy.
In addition, {\AOMP} was shown to enjoy less restrictive exact recovery guarantees for noise-free signals \cite{Karahanoglu:ImprovingAOMP} than MMP.

Both being based on exploring multiple candidates via search tree structures, {\AOMP} and MMP belong to the same family of algorithms, where {\AOMP} may be seen as a more sophisticated version.
In addition, {\AOMP} is not only an earlier proposal but it is also possible to obtain MMP from {\AOMP} via simplification of the path selection strategy.
The evaluation of MMP cannot be considered complete without being compared to a closely related proposal which shows strong empirical potential.
Consequently, we believe MMP should have been compared to {\AOMP} in terms of both algorithmic relations and empirical performance in the first place.
That neither the algorithmic relations nor empirical comparison of MMP and {\AOMP} is addressed in \cite{MMP2014} has the potential to mislead the reader and should be considered as a vital lack for fair evaluation of MMP.
Though this comment is not meant for technical details, we present a short empirical comparison of {\AOMP} and MMP below to address this issue.

\subsection{Empirical Comparison of MMP and {\AOMP}}
\label{sec:Comparison}

We employ the setup in \cite{Karahanoglu:ImprovingAOMP} for empirical comparison.
Let $\x\in{\mathbb{R}}^{N}$ be a $K$-sparse signal (i.e. having only $K$ nonzero elements)
\footnote{The definitions of $K$ and $M$ are different than the previous section. This choice is made on purpose in order to preserve consistency with the corresponding publications \cite{Cotter:TSBOMP}, \cite{Karahanoglu:AOMPfull}, and \cite{Karahanoglu:ImprovingAOMP}.}.
The observation model is $\y=\mathbf{\Phi}\x$ where $\mathbf{\Phi}\in{\mathbb{R}}^{M\times{N}}$ is the observation matrix and $N>M>K$.
We select $N = 256$, $M=100$ and $K\in[10,50]$.
For each $K$, the test is repeated over 500 randomly generated sparse vectors and Gaussian observation matrices.
We set $I=3$ and $B=2$ for {\AOMP}.
MMP-DF is chosen for comparison, since it is referred to as the practical one in \cite{MMP2014}.
The branching factor of MMP-DF is set to $L=6$ as in \cite{MMP2014}. We allow a maximum of 200 paths for both algorithms.
The average normalized mean-squared-error (ANMSE) is defined as
\begin{equation}\label{Eq:ANMSE}
    \text{ANMSE} = \frac{1}{500} \sum_{i=1}^{500}{\frac{\|\x_i - \hat{\x}_i\|_2^2}{\|\x_i\|_2^2}}
\end{equation}
where $\hat{\x}_i$ is the recovery of the $i$th test vector $\x_i$.

We demonstrate two different termination criteria.
{\AOMPK} and {\MMPKDF} limit the number of nonzero elements by the true $K$.
{\AOMPe} and {\MMPeDF}\footnote{Application of this termination criterion to MMP is new here, and improves the recovery accuracy significantly over \cite{MMP2014}.} terminate when $\|\res\|_2<10^{-6}\|\y\|_2$, where $\res$ is the residue from $\y$.
Practically, the number of nonzero elements is limited to $\Kmax>K$, which is set to $55$ here.
{\AOMPK} uses the multiplicative cost model \cite{Karahanoglu:AOMPfull} with  $\alpha_{\text{Mul}}=0.8$, while {\AOMPe} employs the adaptive-multiplicative cost model \cite{Karahanoglu:ImprovingAOMP} with $\alpha_{\text{AMul}}=0.97$.

\begin{figure}[!t]
 \centering
\includegraphics[width=\linewidth]{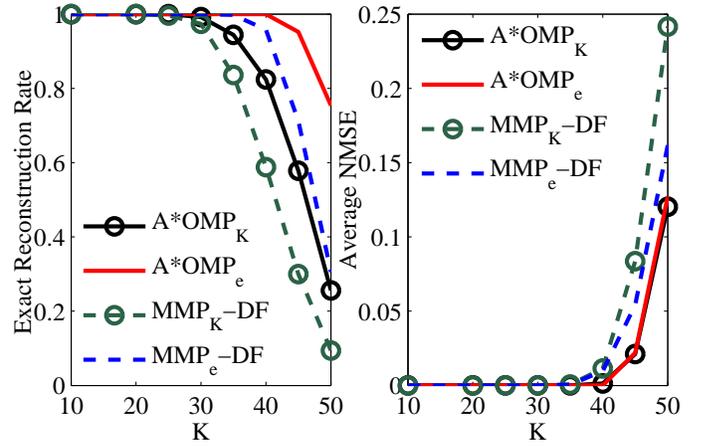}
\caption{Recovery results for the Gaussian sparse signals.}
\label{fig:gauss}
\end{figure}

\begin{figure}[!t]
\centering
\includegraphics[width=\linewidth]{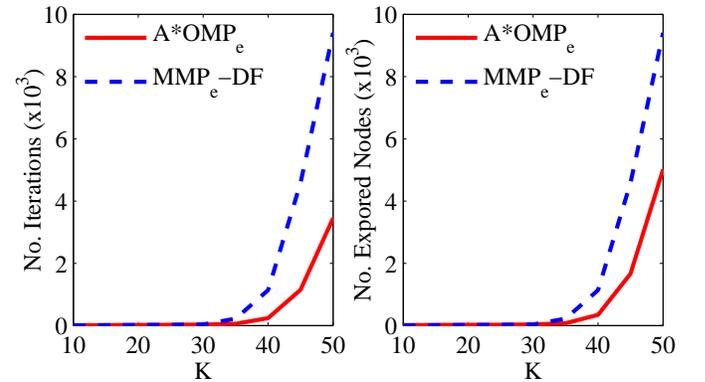}
\caption{Average number of iterations and average number of explored nodes for the Gaussian sparse signals. Note that the numbers should be multiplied with $10^3$ to get the exact figures.}
\label{fig:gauss2}
\end{figure}

 \figurename~\ref{fig:gauss} depicts the exact recovery rates and ANMSE values obtained with {\AOMP} and MMP. It is evident that limiting the search to $K$ nonzero indices is suboptimal. {\AOMP} significantly outperforms MMP for both termination criteria, while {\AOMPe} is the top performer. \figurename~\ref{fig:gauss2} illustrates the average number of iterations and the average number of explored nodes for {\AOMPe} and {\MMPeDF}. We observe that {\AOMPe} requires significantly fewer iterations and explores fewer nodes than {\MMPeDF} does.

These findings are expected since {\AOMP} employs an adaptive path selection mechanism, while MMP simply follows a predefined order. It is clear that the sophisticated path selection mechanism of {\AOMP} is indeed able to guide the search to the true solution with higher probability than MMP, and achieves this by evaluating fewer candidate solutions.

\begin{figure}[!t]
\centering
\includegraphics[width=0.6\linewidth]{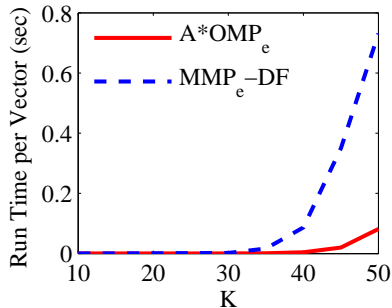}
\caption{Average run times for the Gaussian sparse signals.}
\label{fig:gauss3}
\end{figure}

Another interesting factor to investigate is the average run times. It is important to note that {\AOMPe} tests are conducted using the optimized AStarOMP software\footnote{Available at http://myweb.sabanciuniv.edu/karahanoglu/research/.} developed by the authors, while MMP tests are run in MATLAB\footnote{AStarOMP is based on a trie structure which not only keeps the tree size at minimum, but also provides optimal tree modification routines. Unfortunately, the authors are unaware of any similar implementations of MMP, and developing one is clearly beyond the scope of this comment.}. While a comparison of run times is given in \figurename~\ref{fig:gauss3}, the numbers are not directly comparable due to these implementation details. However, the figure still indicates the significant advantages of using the optimized AStarOMP software.

\section{Validity of Theoretical Analysis}
\label{sec:Theory}

Theoretical performance of MMP has been analysed in \cite{MMP2014}.
This analysis provides exact recovery guarantees based on the restricted isometry property (RIP) for both noise-free and noisy recovery of sparse signals.
Though these results are applicable to the ``theoretical definition'' of the algorithm, they do ignore the pruning strategies which the authors of \cite{MMP2014} incorporate in practice.
As the authors also acknowledge, the computational complexity of MMP is still burdensome without these pruning strategies.
Though such limitations may be accepted, their relations to the theory should be clarified in order to prevent misinterpretation of the analytical findings.
Therefore, the Theorems 3.9 and 4.5 should be considered together with the pruning strategies for their practical implications.
Otherwise, these results may lead to improper observations when compared to other algorithms.
This may be addressed by adding the pruning strategies as assumptions to these theorems as done for {\AOMP} in \cite{Karahanoglu:ImprovingAOMP}.

In addition, the exact recovery guarantees \cite{Karahanoglu:ImprovingAOMP}  developed in \cite{Karahanoglu:ImprovingAOMP} for {\AOMP}  are also valid for MMP.
This analysis provides a looser (RIP)-based exact recovery condition for noise-free recovery of sparse signals than the one in \cite{MMP2014}.
To be concrete, we present the following lemma:

\begin{lemma}
According to the Theorem 2 of \cite{Karahanoglu:ImprovingAOMP}, MMP recovers $K$-sparse signals exactly from measurements $\y=\mathbf{\Phi}\x$  under appropriate pruning assumptions if $\mathbf{\Phi}$ satisfies RIP with the restricted isometry constant (RIC)
\begin{equation}
\delta_{K+L} < \frac{\sqrt{L}}{\sqrt{K}+\sqrt{L}}.
\label{Eq:AOMPK_res1}
\end{equation}

This condition is better than the one in \cite{MMP2014}, since the RIC bound in (\ref{Eq:AOMPK_res1}) is smaller than the following RIC bound from Theorem 3.9 of \cite{MMP2014}:
\begin{equation}
\delta_{K+L} < \frac{\sqrt{L}}{\sqrt{K}+2\sqrt{L}}.
\label{Eq:AOMPK_res2}
\end{equation}
\end{lemma}

\bibliographystyle{IEEEbib}
\bibliography{IEEEAbrv,CommentsOnMMP}

%\begin{IEEEbiography}[{\includegraphics[width=1in,height=1.25in,clip,keepaspectratio]{Photo_Karahanoglu.eps}}]
%{Nazim Burak Karahanoglu} is a senior researcher at the Informatics and Information Security Research Center of the Scientific and Technological Research Council of Turkey (TUBITAK BILGEM) in Kocaeli, Turkey. He received his B.S. degree in Electrical and Electronics Engineering from METU, Ankara  in 2003, M.S. degree in Computational Engineering from the Friedrich-Alexander University of Erlangen-Nuremberg, Germany in 2006 and Ph.D. degree in Electronics Engineering from Sabanci University, Istanbul in 2013. He has been with TUBITAK since 2008.  His research interests include compressed sensing, sparse signal recovery, and sonar signal processing.
%\end{IEEEbiography}
%\begin{IEEEbiography}[{\includegraphics[width=1in,height=1.25in,clip,keepaspectratio]{Photo_Erdogan.eps}}]
%{Hakan Erdogan} is an assistant professor at Sabanci University in Istanbul, Turkey. He received his B.S. degree in Electrical Engineering and Mathematics in 1993 from METU, Ankara and his M.S. and Ph.D. degrees in Electrical Engineering: Systems from the University of Michigan, Ann Arbor in 1995 and 1999 respectively. He was with the Human Language Technologies group at IBM T.J. Watson Research Center, NY between 1999 and 2002. He has been with Sabanci University since 2002. His research interests are in developing and applying probabilistic methods and algorithms for multimedia information extraction.
%\end{IEEEbiography}

\begin{biographynophoto}{Nazim Burak Karahanoglu} is a senior researcher at the Informatics and Information Security Research Center of the Scientific and Technological Research Council of Turkey (TUBITAK BILGEM) in Kocaeli, Turkey. He received his B.S. degree in Electrical and Electronics Engineering from METU, Ankara  in 2003, M.S. degree in Computational Engineering from the Friedrich-Alexander University of Erlangen-Nuremberg, Germany in 2006 and Ph.D. degree in Electronics Engineering from Sabanci University, Istanbul in 2013. He has been with TUBITAK since 2008.  His research interests include compressed sensing, sparse signal recovery, and sonar signal processing.
\end{biographynophoto}

\begin{biographynophoto}{Hakan Erdogan} is a professor at Sabanci University in Istanbul, Turkey. He received his B.S. degree in Electrical Engineering and Mathematics in 1993 from METU, Ankara and his M.S. and Ph.D. degrees in Electrical Engineering: Systems from the University of Michigan, Ann Arbor in 1995 and 1999 respectively. He was with the Human Language Technologies group at IBM T.J. Watson Research Center, NY between 1999 and 2002. He has been with Sabanci University since 2002. His research interests are in developing and applying probabilistic methods and algorithms for multimedia information extraction.
\end{biographynophoto}
\vfill

\end{document}